\documentclass[%
reprint,
superscriptaddress,
 amsmath,amssymb,
 aps,
]{revtex4-1}

\usepackage{graphicx}
\usepackage{float}
\usepackage{braket}
\usepackage{subcaption}
\usepackage{color, ulem}
\usepackage{gensymb}

\begin{document}

\title{Critical role of device geometry for the phase diagram of twisted bilayer graphene}

\author{Zachary A. H. Goodwin}
\affiliation{Departments of Materials and Physics and the Thomas Young Centre for Theory and Simulation of Materials, Imperial College London, South Kensington Campus, London SW7 2AZ, UK\\}
\author{Valerio Vitale}
\affiliation{Departments of Materials and Physics and the Thomas Young Centre for Theory and Simulation of Materials, Imperial College London, South Kensington Campus, London SW7 2AZ, UK\\}
\author{Fabiano Corsetti}
\affiliation{Departments of Materials and Physics and the Thomas Young Centre for Theory and Simulation of Materials, Imperial College London, South Kensington Campus, London SW7 2AZ, UK\\}
\author{Dmitri K. Efetov}
\affiliation{ICFO - Institut de Ciencies Fotoniques, The Barcelona Institute of Science and Technology, Castelldefels, Barcelona, 08860, Spain\\}
\author{Arash A. Mostofi}
\affiliation{Departments of Materials and Physics and the Thomas Young Centre for Theory and Simulation of Materials, Imperial College London, South Kensington Campus, London SW7 2AZ, UK\\}
\author{Johannes Lischner}
\affiliation{Departments of Materials and Physics and the Thomas Young Centre for Theory and Simulation of Materials, Imperial College London, South Kensington Campus, London SW7 2AZ, UK\\}

\date{\today}

\begin{abstract}
The effective interaction between electrons in two-dimensional materials can be modified by their environment, enabling control of electronic correlations and phases. Here, we study the dependence of electronic correlations in twisted bilayer graphene (tBLG) on the separation to the metallic gate(s) in two device configurations. Using an atomistic tight-binding model, we determine the Hubbard parameters of the flat bands as a function of gate separation, taking into account the screening from the metallic gate(s), the dielectric spacer layers and the tBLG itself. We determine the critical gate separation at which the Hubbard parameters become smaller than the critical value required for a transition from a correlated insulator state to a (semi-)metallic phase. We show how this critical gate separation depends on twist angle, doping and the device configuration. These calculations may help rationalise the reported differences between recent measurements of tBLG's phase diagram and suggests that correlated insulator states can be screened out in devices with thin dielectric layers. 
\end{abstract}

\maketitle

\section{Introduction}

Twisted bilayer graphene (tBLG) has emerged as a highly tunable platform (through twist angle, hydrostatic pressure, and doping) for studying the behaviour of correlated electrons in two dimensions~\cite{NAT_I,NAT_S,TSTBLG,SOM,NAT_SS,NAT_MEI,NAT_CO,PDTBLG}. Several groups have reported the experimental observation of correlated insulator states in both doped and undoped tBLG near the magic twist angle~\cite{NAT_I,NAT_S,TSTBLG,SOM}. Despite much theoretical work, no consensus has yet been reached regarding the microscopic mechanism of these phases~\cite{IMACP,KVB,SCDID,EPRG,SCPKV,KL,CSD,SCHFC,WC, EPC,OMIB,NCIS,PMS,PCIS,TCSSV,US,EE}. Superconductivity is also found at low temperatures~\cite{NAT_S,TSTBLG,SOM}, but neither the nature of the pairing interaction nor the symmetry of the superconducting order parameter have been determined.

Importantly, significant differences in the measured phase diagrams have been reported~\cite{NAT_I,NAT_S,TSTBLG,SOM}: some groups have found correlated insulator states at doping levels where other groups find semi-metallic or metallic behaviour~\cite{NAT_I,NAT_S,TSTBLG,SOM}. Also, different numbers of superconducting states as a function of doping have been reported~\cite{NAT_S,TSTBLG,SOM}. It has been suggested that these differences are a consequence of the varying degree of twist-angle homogeneity in the tBLG samples~\cite{SOM}. 

Another potential origin of the variations in the observed phase diagram are differences in the device setups employed in the experiments. For example, Cao~\textit{et al.}~\cite{NAT_I,NAT_S} used devices in which the tBLG is encapsulated by hexagonal boron-nitride (hBN) slabs of 10-30~nm thickness with gold gates on either side. In contrast, Yankowitz~\textit{et al.}~\cite{TSTBLG} used thicker hBN slabs (30-60~nm) that are sandwiched between two graphite gates, and the device of Lu~\textit{et al.}~\cite{SOM} only had a single graphite gate separated from the tBLG by a hBN layer of $\sim10$~nm thickness. 

The device setup is important because the dielectric environment in which tBLG is situated will have a strong influence on the screened interaction between the electrons in tBLG~\cite{CCRPA}, which will in turn affect the phase diagram. For gated tBLG devices, the potential induced by an electron in the tBLG is reduced by the image charge in the metallic gate(s). The strength of the resulting effective interaction between electrons in the tBLG is determined by the separation of the metallic gate(s) from the tBLG which can be experimentally controlled via the thickness of a dielectric spacer layer (typically hBN). 

In this article, we investigate the dependence of electron correlations in tBLG devices on the distance between tBLG and the metallic gate(s) (or equivalently, the thickness of the hBN layer). Starting from an atomistic tight-binding model, we construct Wannier functions of the flat bands and calculate the corresponding Hubbard parameters taking into account screening from the metallic gates, the hBN and the tBLG itself. We find that the Hubbard parameters depend sensitively on the gate distance enabling precise control of electron correlations via device geometry. We calculate the critical gate distances at which the correlated insulator states disappear and predict detailed phase diagrams as function of doping. We also discuss the interplay of device geometry and superconductivity and hypothesize that small gate separations should weaken correlated insulator states, but strengthen superconducting phases.

\begin{figure}
\centering
\includegraphics[width=1\linewidth]{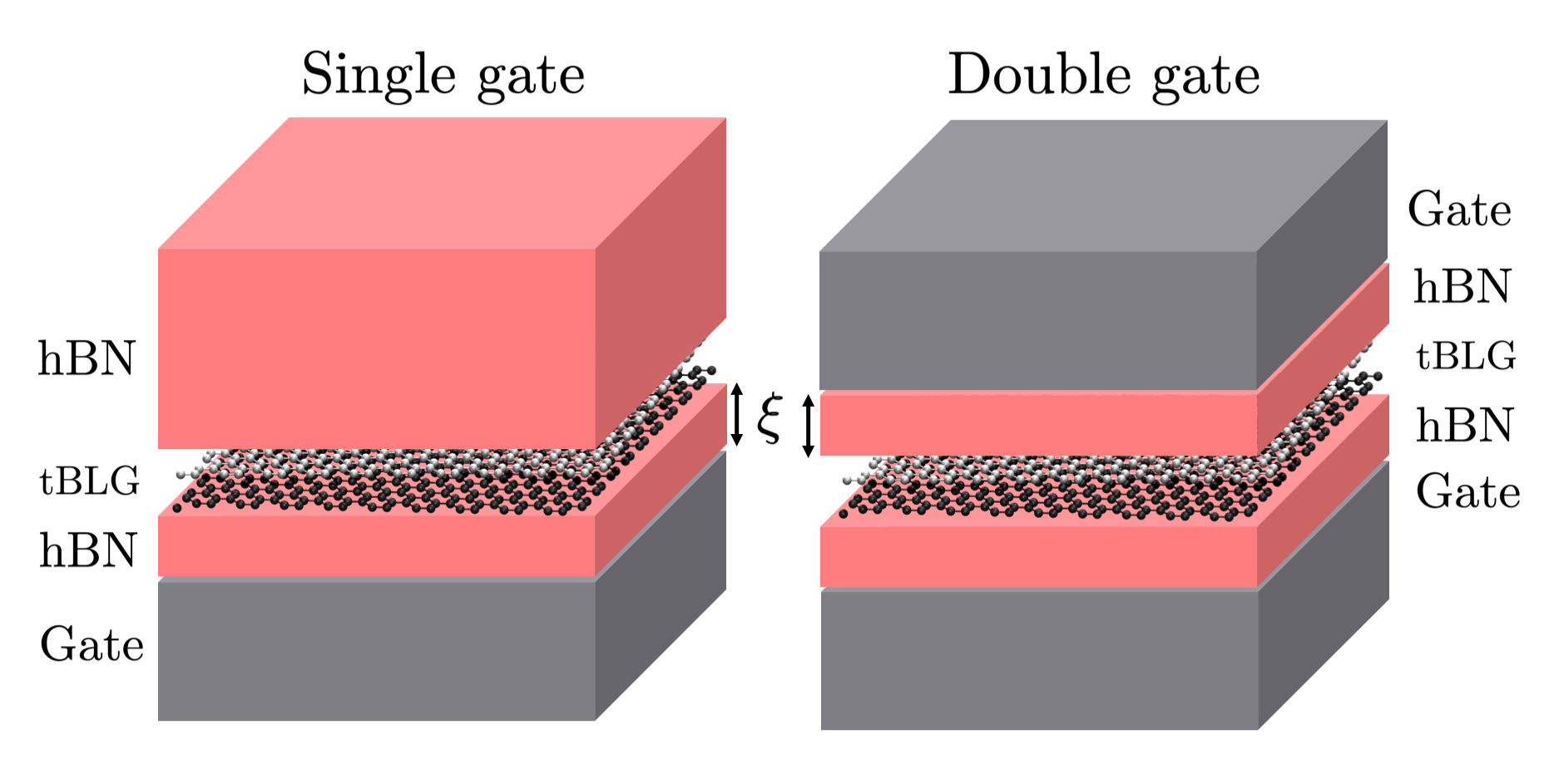}
\caption{Twisted bilayer graphene devices with a single gate (left) and two gates (right). The separation between the tBLG and the gate(s) is denoted by $\xi$. For the double-gated device, the separation is set to be equal on either side.}
\label{fig:EXP_DEV}
\end{figure}

\section{Methods} 

To model the electronic structure of tBLG near the magic angle, we employ the techniques described in detail in Refs.~\citenum{PHD_1} and ~\citenum{PHD_2}. For convenience, we summarize briefly the method here. 
To obtain the electronic band structure of tBLG, we carry out atomistic tight-binding calculations using the parameters of Ref.~\citenum{NSCS}. Next, Wannier functions $w_i(\mathbf{r})$ (with $i$ denoting both the unit cell and the band-like index of the Wannier function) of the flat bands are generated~\cite{MLWO,MLWF,W90vT,SMLWF,PHD_1,PHD_2}. The extended Hubbard parameters of the flat band electrons are obtained by evaluating 
\begin{equation}
V_{ij} = \iint d\textbf{r}d\textbf{r}^{\prime}|w_{i}(\textbf{r}^{\prime})|^{2}W(\textbf{r} - \textbf{r}^{\prime})|w_{j}(\textbf{r})|^{2},
\label{GHI}
\end{equation}
\noindent where $W(\mathbf{r}-\mathbf{r}')$ denotes the screened interaction between electrons, which is determined by screening processes inside tBLG (internal screening), but also has contributions from the environment. 

In a typical device configuration, the tBLG is encapsulated by a dielectric substrate [usually hexagonal boron nitride (hBN)] which separates it from the metallic gate(s). Here, we investigate two device types: (i) tBLG with a bottom gate only (i.e., gate/hBN/tBLG/hBN, where it is assumed that the hBN on top is very thick and can be approximated by a semi-infinite dielectric, while the hBN layer that separates the tBLG and the gate has a thickness $\xi$, see left panel of Fig.~\ref{fig:EXP_DEV}); and (ii) tBLG with both a top and a bottom gate (i.e., gate/hBN/tBLG/hBN/gate, where both hBN layers have the same thickness $\xi$, see right panel of Fig.~\ref{fig:EXP_DEV}). The gates are modelled as semi-infinite ideal metals.

For the single-gate device, the electrostatic image charge interaction results in a screened interaction of the form 
\begin{equation}
W(\textbf{r} - \textbf{r}^{\prime}) = \dfrac{e^{2}}{4\pi\epsilon_{0}\epsilon_{\textrm{r}}}\Bigg[\dfrac{1}{|\textbf{r} - \textbf{r}^{\prime}|} - \dfrac{1}{|2\xi \textbf{e}_{z} + \textbf{r} - \textbf{r}^{\prime}|}\Bigg],
\label{SMG}
\end{equation}
\noindent where $e$ is the electron charge, $\epsilon_0$ is the permittivity of free space, and $\epsilon_\textrm{r}$ denotes the dielectric constant due to screening from the tBLG and the hBN. We have assumed that the tBLG lies in the $xy$-plane ($\textbf{e}_z$ denotes the unit vector pointing in the $z$-direction) and that $\xi$ corresponds to the thickness of hBN (distance from the metallic gate to the bottom of the tBLG). In principle, the internal screening from the tBLG leads to a complicated wave-vector dependent dielectric function, but Goodwin \textit{et al.}~\cite{PHD_2} have previously shown that---near the magic angle---the Hubbard parameters within the constrained random phase approximation (cRPA) can be accurately reproduced using a bare Coulomb interaction divided by the cRPA dielectric constant of a decoupled bilayer graphene (8.86)~\cite{PHD_2}. To describe the screening by the dielectric substrate we add to this the bulk dielectric constant of hBN (3.9)~\cite{GhBN} and subtract one to avoid double-counting the dielectric constant of free space resulting in $\epsilon_{\rm r}=11.76$. 
For the double-gate device, the screened interaction is given by~\cite{MGS}
\begin{equation}
W(\textbf{r} -\textbf{r}^{\prime}) = \dfrac{e^{2}}{4\pi\epsilon_{\textrm{r}}\epsilon_{0}}\sum_{n=-\infty}^{+\infty}\dfrac{(-1)^{n}}{\sqrt{|\textbf{r} -\textbf{r}^{\prime}|^{2} + (2\xi n)^{2}}}.
\label{WMG}
\end{equation}
\noindent For in-plane distances much smaller than $\xi$, this screened interaction reduces to the bare Coulomb interaction divided by $\epsilon_{\rm r}$. For in-plane distances much larger than $\xi$, the screened interaction can be expressed as $W(r) = e^{2}e^{-\pi r/\xi}/(2\pi\epsilon_{\rm r}\epsilon_{0}\sqrt{r\xi})$~\cite{MGS}. Here we assume that in the image-potential the tBLG resides in the $z = 0$ plane, which again means $\xi$ corresponds to the thickness of the hBN spacer.

\begin{figure*}[t!]
\begin{subfigure}{0.49\textwidth}
  \centering
  \includegraphics[width=1\linewidth]{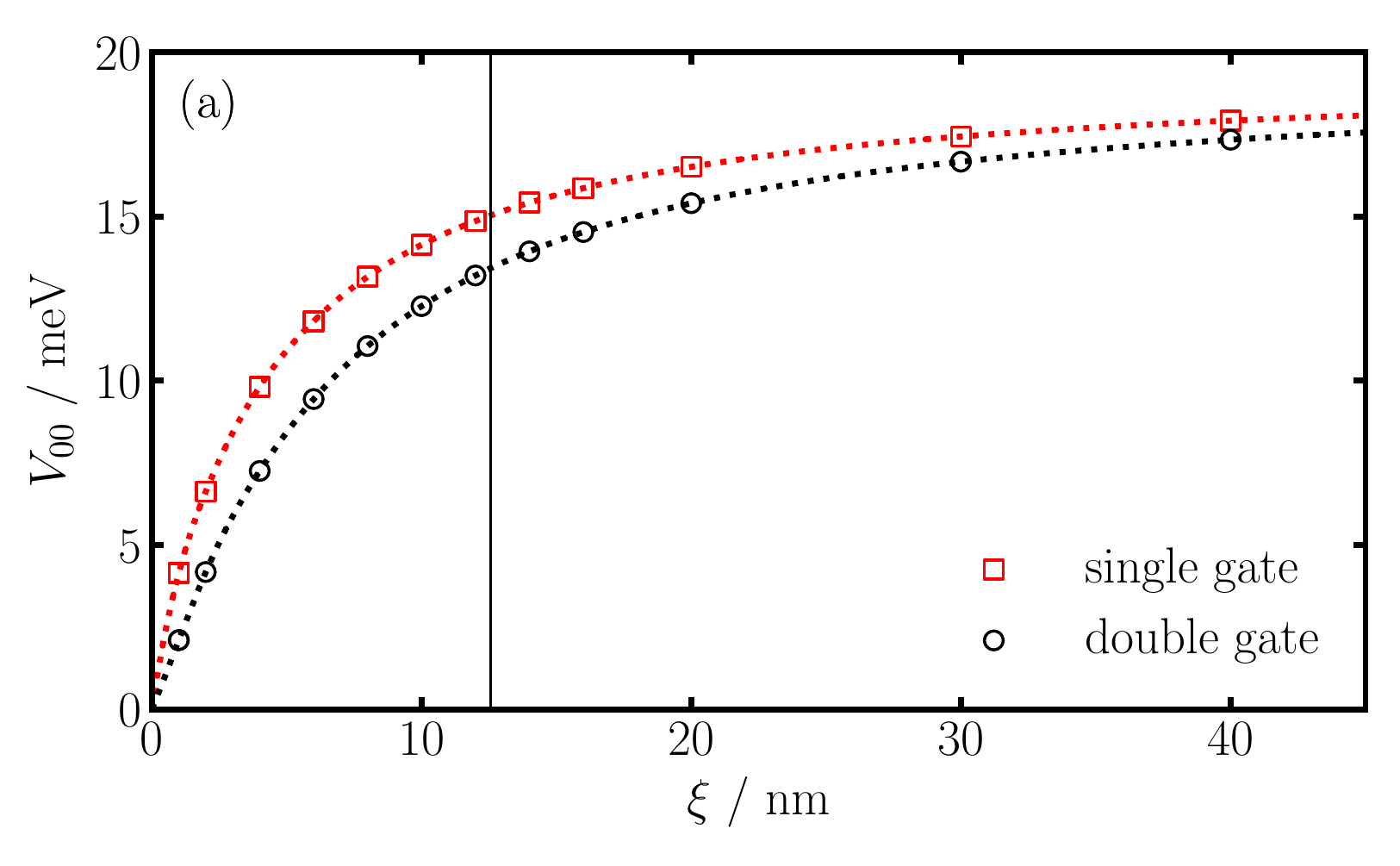}
\end{subfigure}
\begin{subfigure}{0.49\textwidth}
  \centering
  \includegraphics[width=1\linewidth]{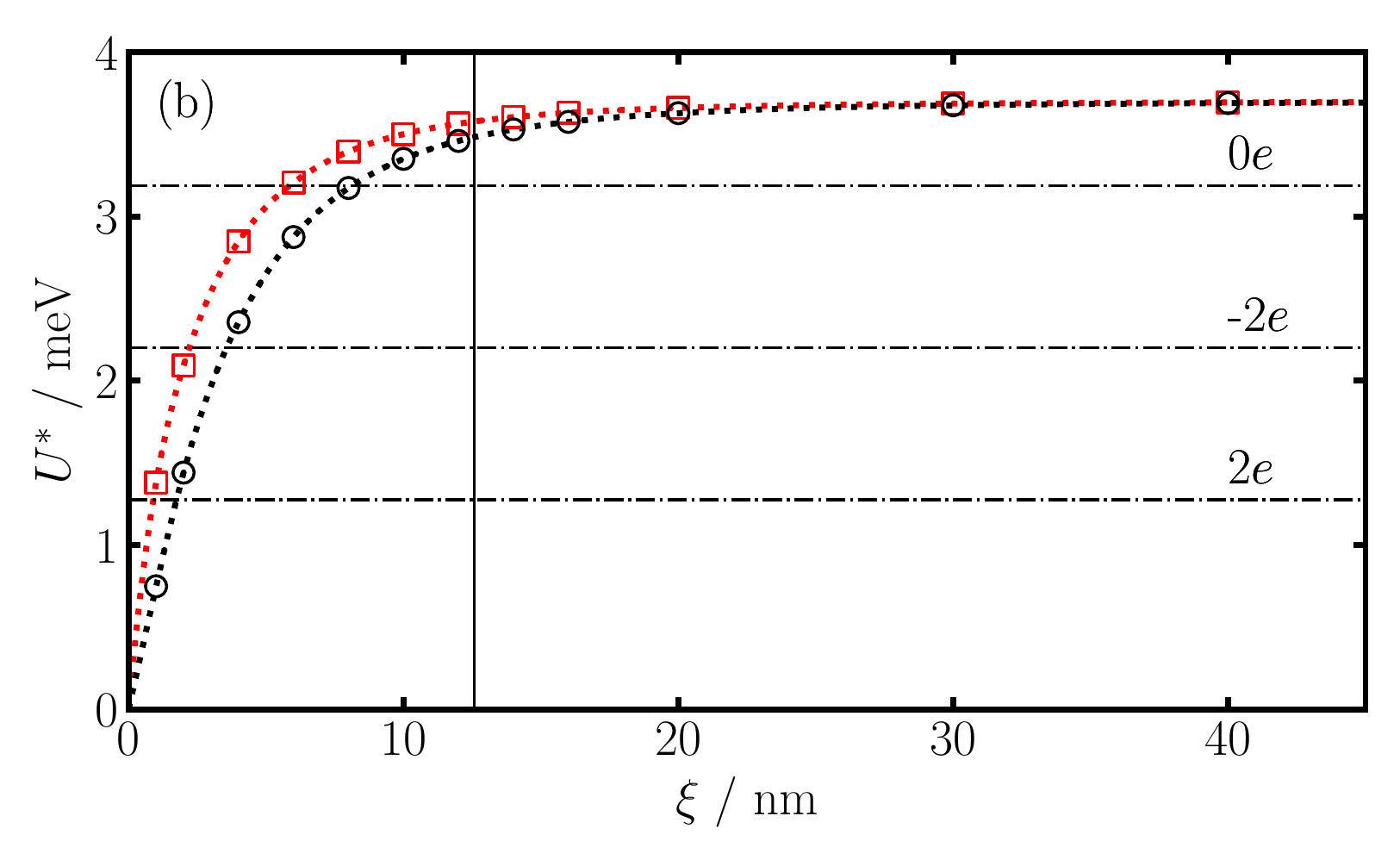}
\end{subfigure}
\caption{(a) On-site Hubbard parameter $V_{00}$ as a function of distance $\xi$ between tBLG and the metallic gate(s) for a twist angle of 1.12$\degree$ for the single-gate and double-gate device (see Fig.~\ref{fig:EXP_DEV} for device geometries). (b) Long-range corrected on-site Hubbard parameter $U^*$ as a function of $\xi$ for a twist angle of 1.12$\degree$ for the single-gate and double-gate device. The horizontal dotted-dashed lines correspond to the critical values of $U^*$ from Ref.~\citenum{LK_CH}. Dotted lines through the data points correspond to fits that are used to extract critical gate separations in Fig.~\ref{Xic_SMG}. The solid vertical line denotes to the length of the moir\'e unit cell.}
\label{U_Us}
\end{figure*}

\section{Results and Discussion} 

Figure~\ref{U_Us}(a) shows the on-site Hubbard parameter $V_{00}$ of tBLG as a function of the gate separation $\xi$ for a twist angle of 1.12$\degree$, for both single-gate and double-gate device configurations (other twist angles are shown in Appendix~\ref{AA}). The largest values of $V_{00}$ are obtained for large gate separations, and there is a substantial reduction in $V_{00}$ as the gate separation decreases below the moir\'e length (indicated by the vertical line in Fig.~\ref{U_Us}). This can be understood from Eqs.~\eqref{SMG} and \eqref{WMG} where the image charge contribution arising from the induced charge density at the surface of the gate reduces the Coulomb interaction between electrons in the tBLG. In the case of the single-gate device, when $\xi$ is small, the electron in the tBLG and its image charge effectively interact with other electrons via a weak dipolar potential (instead of the usual monopole charge-charge interaction). For the double-gate setup, the image charges in both gates give rise to an exponentially screened interaction between electrons in the tBLG. The results for $V_{00}$ in the two different device configurations are qualitatively similar, but the Hubbard parameters are somewhat smaller for the double-gate setup since both of the gates contribute to the screening.

In the Hubbard model~\cite{HUB_I,HUB_II}, electron-electron interactions are assumed to be short-ranged and the strength of electron correlations is usually measured by the ratio of the on-site Hubbard parameter $V_{00}$ and the nearest-neighbor hopping parameter $t$. A system with long-ranged electron-electron interactions can be mapped onto an effective Hubbard model with an on-site Hubbard parameter $U^* = V_{00} - V_{01}$ (reflecting the energy required to hop from an empty Wannier orbital to an occupied neighbor site), where $V_{01}$ is the interaction between Wannier functions centred on neighbouring AB/BA regions (see Fig.~\ref{U_Us_THETA} of Appendix~\ref{AA} for $V_{01}$ as a function of $\xi$)~\cite{OHP,PHD_1,PHD_2}. Fig.~\ref{U_Us}(b) shows $U^*$ for tBLG as function of $\xi$ for a twist angle of 1.12$\degree$ for both device configurations (results for other twist angles are shown in Appendix~\ref{AA}). It can be seen that $U^*$ is significantly smaller than $V_{00}$, indicating that interactions between neighboring Wannier functions play an important role even in the presence of metallic gates~\cite{PHD_1,PHD_2}. This is expected as there is significant overlap between lobes of neighboring Wannier functions~\cite{MLWO,PHD_1}. Similarly to $V_{00}$, $U^*$ approaches a constant value in the limit of distant gates, but does so more rapidly once $\xi$ becomes larger than the moir\'e length. Moreover, $U^*$ exhibits a significantly sharper reduction as the  distance decreases. Naively, one might expect that the presence of gates should lead to an \textit{increase} in $U^*$ if the screened interaction is sufficiently short-ranged such that $V_{01}$ is strongly reduced. We find indeed that $V_{01}$ decreases more quickly than $V_{00}$ (see Appendix~\ref{AA}), but this \textit{relative} reduction of $V_{01}$ compared to $V_{00}$ is not sufficient to overcome the large \textit{absolute} reduction of $V_{00}$ [see Fig.~\ref{U_Us}(a)] and, therefore, the overall balance is such that $U^*$ decreases with decreasing $\xi$. 

%ZG: I don't think we need to add anything else here
%\textcolor{red}{[AAM: I wonder whether the first part of the idea being conveyed here could be explained more clearly. I think what is meant here is that the fractional (or, equivalently, percentage) reduction in $V_{01}$ is faster than $V_{00}$. In other words, $V_{01}(\xi)/V_{01}(\infty)$ decreases faster with decreasing $\xi$ than $V_{00}(\xi)/V_{00}(\infty)$? Could we show this in a plot in the Appendix and thereby strengthen the link between the main text and the Appendix?]}

\subsection{Correlated Insulators} 

When $U^*/t$ reaches a critical value, a phase transition from a (semi-)metallic phase to a correlated insulator state is expected. For tBLG, no consensus has yet been reached regarding the nature of the correlated insulator states, nor the corresponding value of the critical $U^*/t$. For Bernal stacked bilayer graphene, Quantum Monte Carlo calculations have found a critical value of $2.2$ for a phase transition to a gapped antiferromagnetic phase~\cite{ABS}. Recently, Klebl and Honerkamp calculated the phase diagram of tBLG using an atomistic RPA approach~\cite{LK_CH}. For undoped tBLG (denoted $0e$), they find a transition from a semi-metal to an antiferromagnetic insulator, and for doping levels corresponding to two extra electrons ($-2e$) or holes ($+2e$) per moir\'e cell, a transition from a metallic phase to a ferromagnetic insulator is predicted. The RPA value for the critical $U^*/t$ in tBLG is smaller than in the Bernal stacked bilayer, and depends both on temperature and doping. Because of the lack of self-energy corrections, the critical $U^*/t$ values from the RPA should be considered as lower bounds~\cite{LK_CH}.

\begin{figure*}[t!]
\includegraphics[width=1\linewidth]{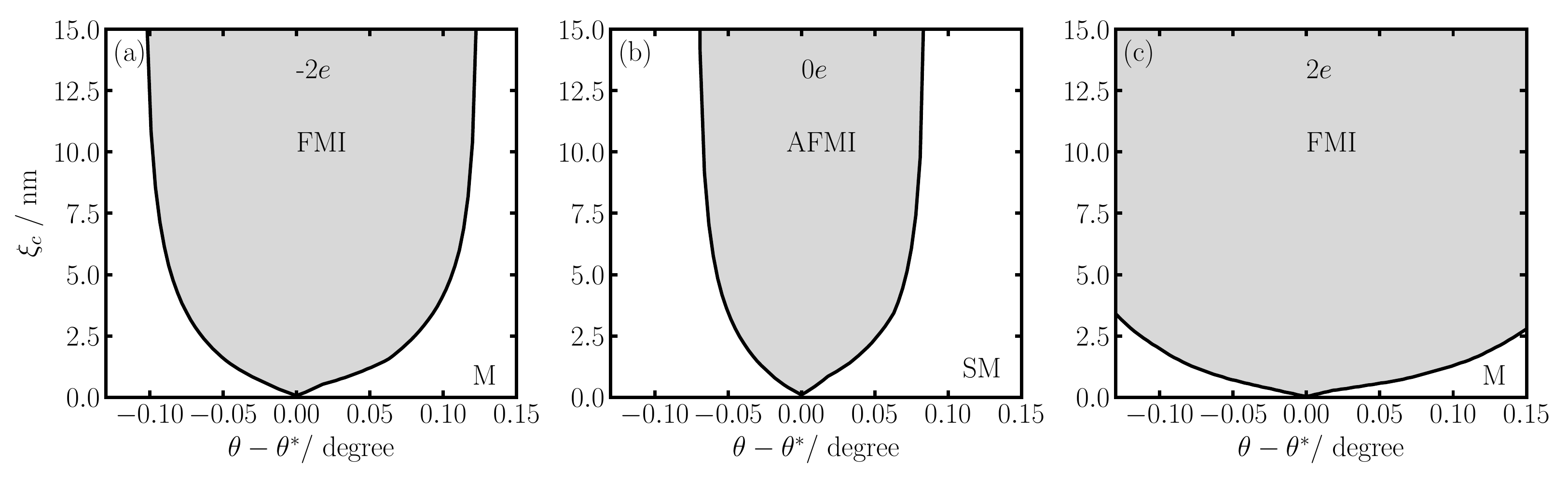}
\caption{Critical value of gate distance as a function of twist angle for (a) two additional electrons per moir\'e unit cell ($-2e$), (b) charge neutrality ($0e$) and (c) two additional holes per moir\'e unit cell ($2e$) for a device with a single gate. The grey regions indicate correlated insulator states (either ferromagnetic insulators (FMI) or anti-ferromagnetic insulators (AFMI)), while the white regions denotes either metallic (M) or semi-metallic (SM) phases.}
\label{Xic_SMG}
\end{figure*}

In Fig.~\ref{U_Us}(b), the critical values of $U^*$ for the cases of $0e$, $-2e$ and $+2e$ doping are indicated by horizontal dotted-dashed lines. Here, we have multiplied the critical $U/t_{\rm G}$ (where $t_{\rm G}$ is the hopping parameter of graphene) values from Klebl and Honerkamp~\cite{LK_CH} at a temperature $T\approx0.3$~K with the hopping parameter $t$ between neighbouring Wannier functions (calculated from the width $\Delta$ of the flat bands in our atomistic tight-binding model using $t=\Delta/6$, which is the relation between bandwidth and hopping in graphene). For the single-gate (double-gate) device, as the gate separation is reduced to $\xi_c=5.86$~nm ($\xi_c=8.14$~nm), $U^*$ crosses the critical value for zero doping, indicating that tBLG would exhibit a semi-metallic phase at zero doping, but correlated insulator states at $-2e$ and $+2e$ doping. At $\xi_c=2.21$~nm ($\xi_c=3.56$~nm), the critical $U^*$ for $-2e$ doping is crossed and finally, at $\xi_c=0.89$~nm ($\xi_c=1.74$~nm) the critical value for $+2e$ doping is reached. For even smaller values of $\xi$, the tBLG is either metallic or semi-metallic at the doping levels considered here. These results demonstrate that the phase diagram of tBLG can be controlled via the thickness of the dielectric substrate that separates the tBLG from the metallic gates and, hence, determines the degree to which the gate is able to screen electron-electron interactions in tBLG. The critical separations for phase transitions depend on the device configuration, with smaller values for single-gate devices because the screening is weaker than in double-gate devices. 

The critical gate separation $\xi_{\rm c}$ also depends on the twist angle. Fig.~\ref{Xic_SMG} shows this dependence as a function of twist angle from the magic angle $\theta^* = 1.18\degree$ for the single-gate device configuration and for three doping levels: $-2e$ (left panel), $0e$ (middle panel) and $+2e$ (right panel) at $T\approx0.3$~K. The equivalent result for the double-gate configuration is shown in Appendix~\ref{AC}. For all doping levels, $\xi_c$ decreases as the magic angle is approached. Close to the magic angle, the hopping approaches zero and extremely small values of $U^*$ must be achieved to reach the critical value of $U^*/t$. This is only possible for very small values of $\xi$. Comparing the three doping levels, we find that $\xi_c$ for the undoped system increases most rapidly away from the magic angle. At twist angles larger than $0.1\degree$ from the magic angle, the undoped system is always metallic and no phase transition to a correlated insulator phase can be induced. For $-2e$ doping, the critical twist angle window is larger than for zero doping. For $+2e$ doping, a critical thickness can be found for all considered twist angles near the magic angle. At larger temperatures, the critical value of $U^*/t$ is smallest at charge neutrality~\cite{LK_CH}, which means that it will require the thinnest hBN slabs to reach the critical value for a phase transition to a (semi-)metallic state. In Appendix~\ref{AD} we show an analogous phase diagrams for both device structures at $T\approx 5~K$. 
\subsection{Superconductivity} 

The thickness $\xi$ of the dielectric spacer layer also influences the stability of the superconducting state which competes with the correlated insulator states discussed above. To bind electrons into Cooper pairs, the effective electron-electron interaction must contain an attractive part $V_{\rm att}$ (this could arise either from electron-phonon coupling, exchange of spin fluctuations, plasmons or any other glue). The total interaction can be expressed as the sum of the bare Coulomb interaction and $V_{\rm att}$. The superconducting transition temperature is approximately given by $T_{\rm c} \propto E_{\rm glue}/k_B \times\exp(-1/(\lambda - \mu^*))$, where $\lambda$ describes the coupling of the electrons to the glue (which has an energy scale $E_{\rm glue}$) and $\mu^*$ is the Coulomb pseudopotential, which describes the repulsion due to the bare Coulomb interaction~\cite{TcS} ($k_B$ is the Boltzmann constant). In the presence of metallic gates, the repulsive bare interaction is reduced by the image charge interaction. As a consequence, $\mu^*$ is also reduced and has a dependence on the gate separation $\xi$. The presence of gates, therefore, should enhance the stability of the superconducting phase and increase the superconducting transition temperature, while reducing the stability of the correlated insulator states. Further calculations are required to quantitatively study the competition between superconductivity and correlated insulator states in the presence of metallic gates.

\section{Summary} 

We have demonstrated the importance of the device geometry for electron correlations and the phase diagram of twisted bilayer graphene. By reducing the separation between the tBLG and the metallic gate(s), the on-site Hubbard parameter can be reduced to a value smaller than the critical value required for a phase transition from a correlated insulator state to a (semi)-metallic state. We have calculated the critical gate-tBLG separation at which correlated insulator states disappear and studied its dependence on twist angle for different doping levels and device configurations. For a fixed twist angle, the phase diagram as function of doping of tBLG depends sensitively on the device geometry which could explain the differences reported by various experimental groups. This opens up the exciting possibility to precisely control electronic phases in tBLG through device engineering. 

\section{Acknowledgements}

We thank D. Kennes, C. Karrasch and A. Khedri for helpful discussions. This work was supported through a studentship in the Centre for Doctoral Training on Theory and Simulation of Materials at Imperial College London funded by the EPSRC (EP/L015579/1). We acknowledge funding from EPSRC grant EP/S025324/1 and the Thomas Young Centre under grant number TYC-101.

\bibliographystyle{apsrev4-1}
\bibliography{CTGS}

\clearpage
\newpage

\onecolumngrid
\appendix

\renewcommand{\theequation}{A\arabic{equation}}
\renewcommand{\thefigure}{A\arabic{figure}}
\setcounter{figure}{0} 
\setcounter{equation}{0} 

\section{Hubbard parameters at different twist angles}
\label{AA}

\begin{figure*}[h]
\begin{subfigure}{0.49\textwidth}
  \centering
  \includegraphics[width=0.9\linewidth]{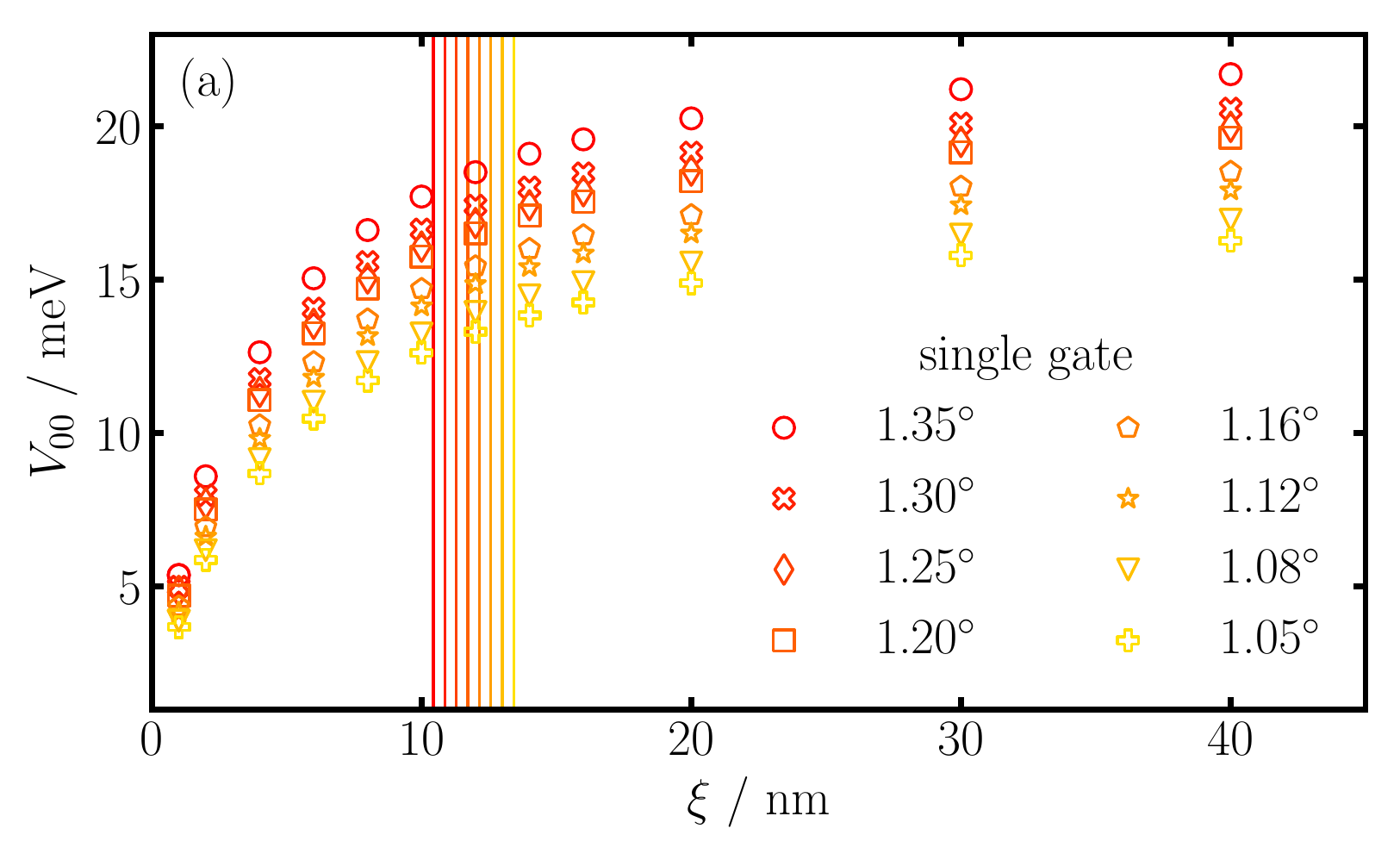}
\end{subfigure}
\begin{subfigure}{0.49\textwidth}
  \centering
  \includegraphics[width=0.9\linewidth]{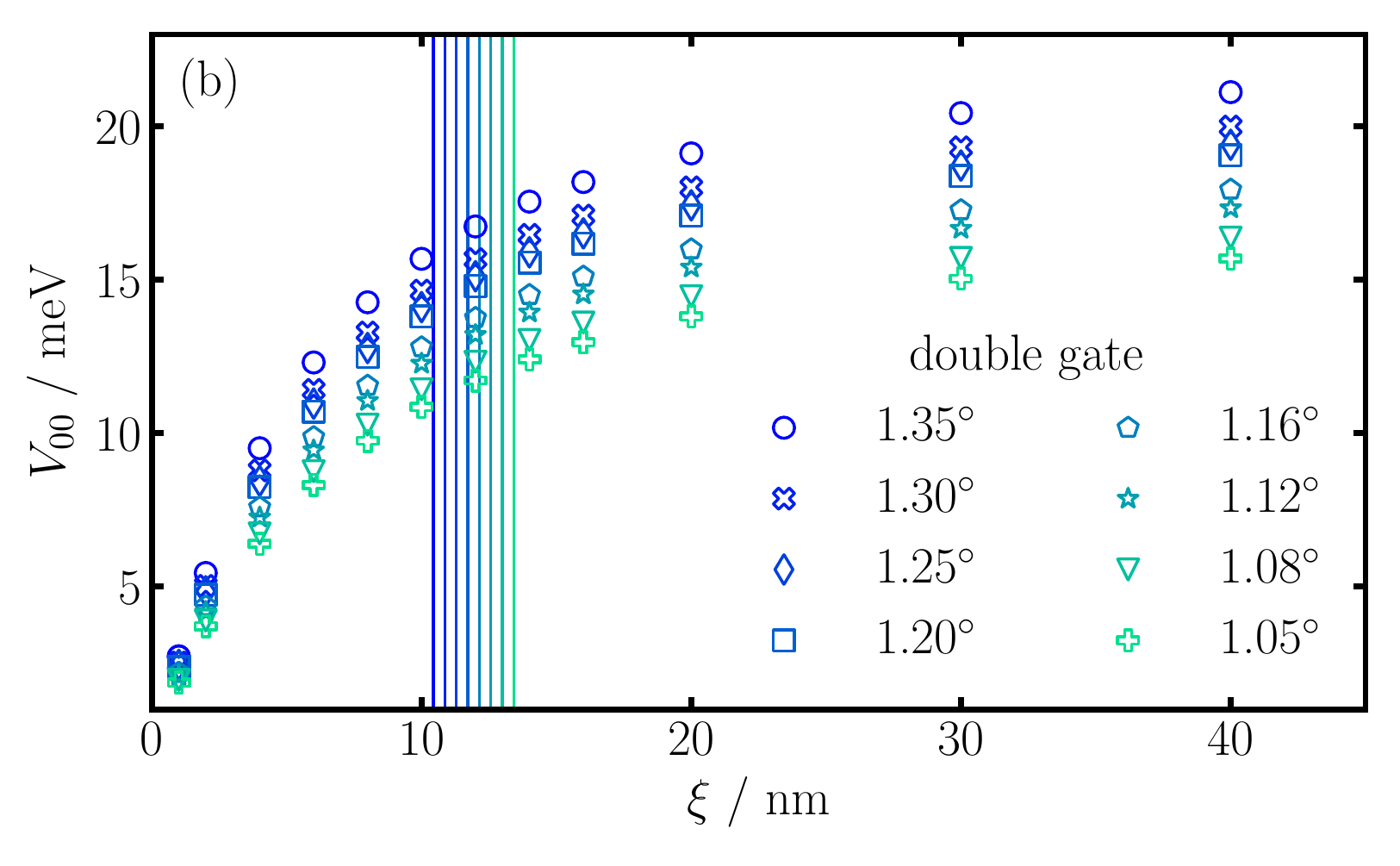}
\end{subfigure}
\begin{subfigure}{0.49\textwidth}
  \centering
  \includegraphics[width=0.9\linewidth]{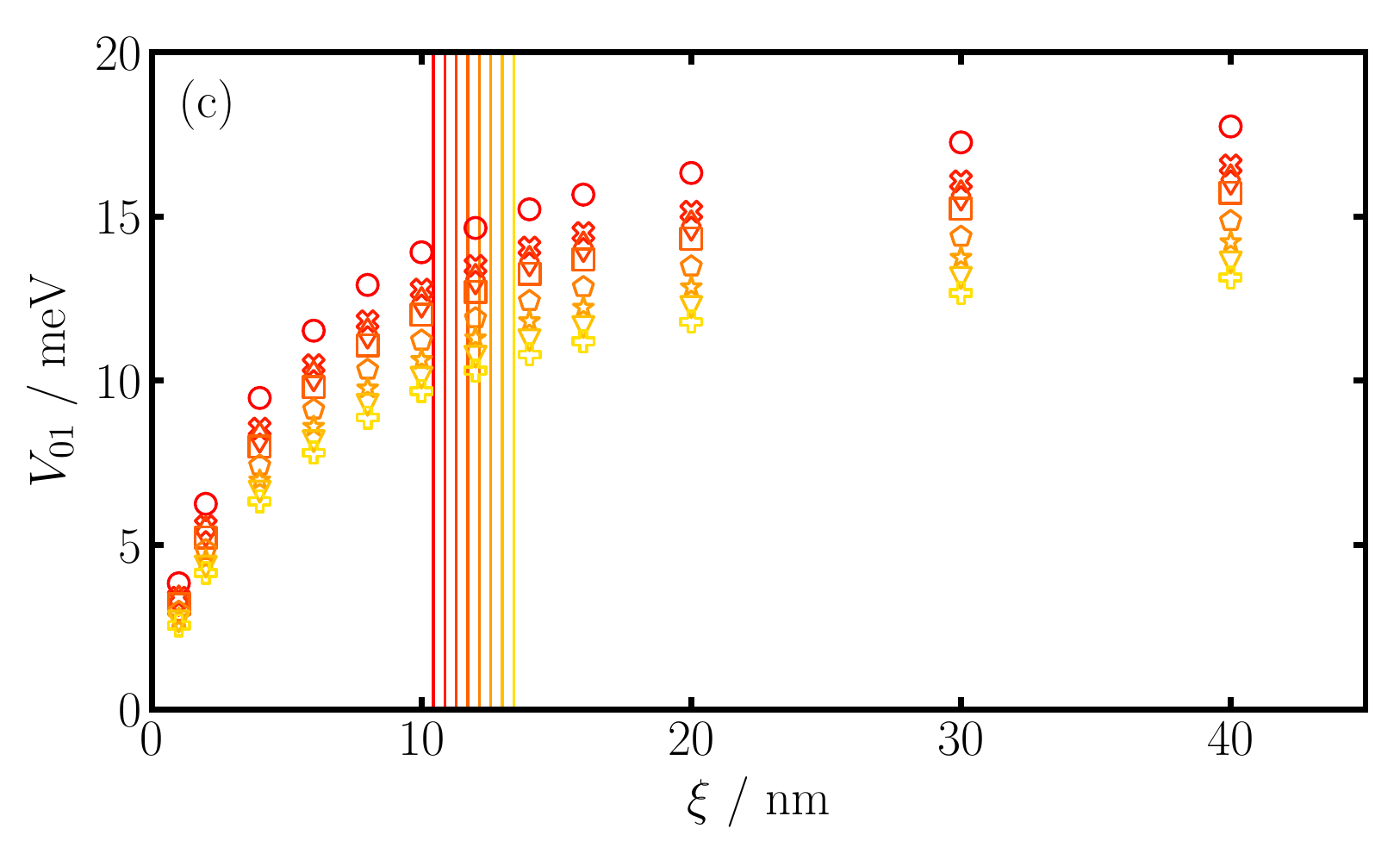}
\end{subfigure}
\begin{subfigure}{0.49\textwidth}
  \centering
  \includegraphics[width=0.9\linewidth]{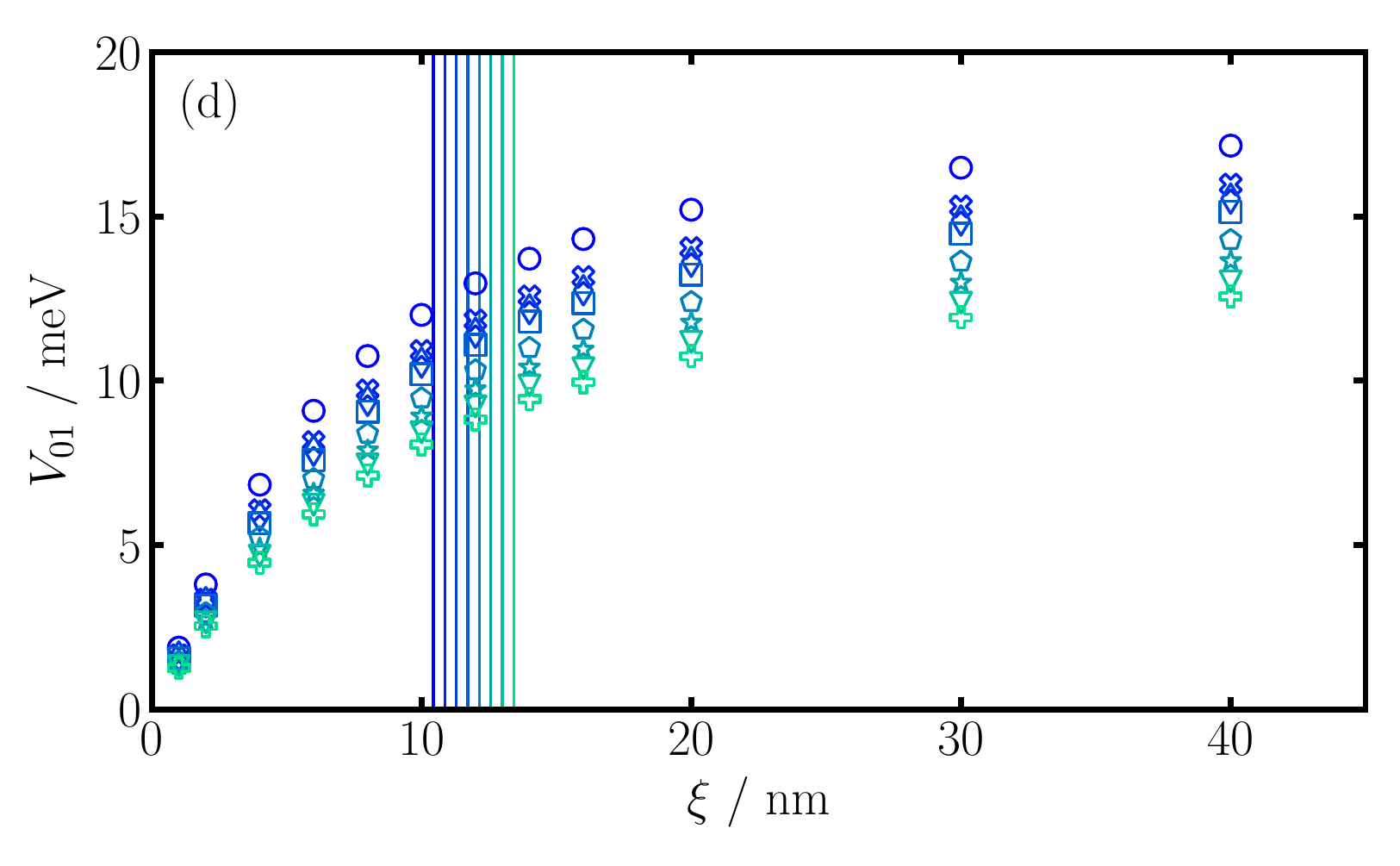}
\end{subfigure}
\begin{subfigure}{0.49\textwidth}
  \centering
  \includegraphics[width=0.9\linewidth]{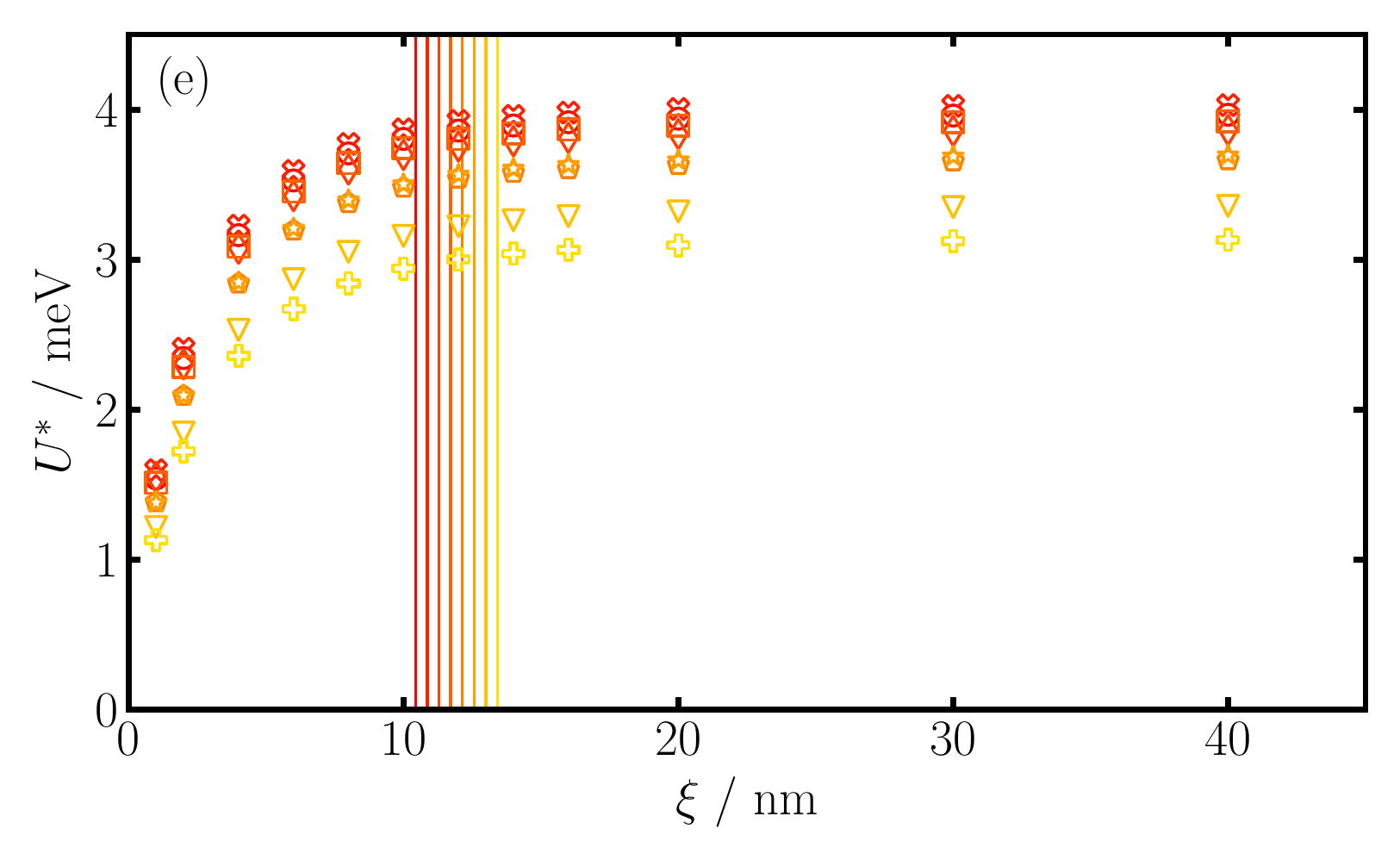}
\end{subfigure}
\begin{subfigure}{0.49\textwidth}
  \centering
  \includegraphics[width=0.9\linewidth]{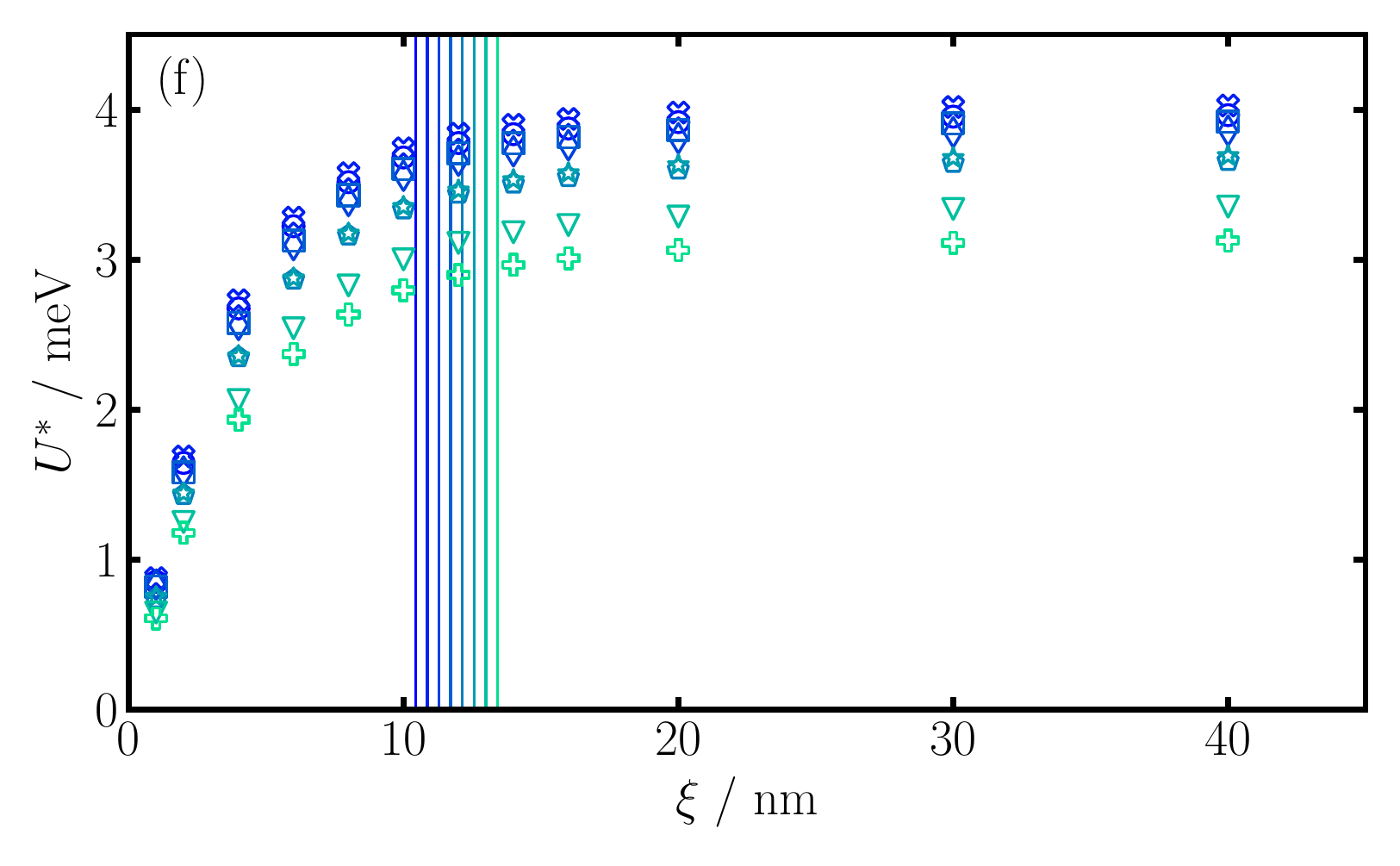}
\end{subfigure}
\caption{Hubbard parameters as a function of the separation $\xi$ of the metallic gate(s) from the tBLG for several twist angles (indicated in the legend). The panels in the left-hand (right-hand) column are for the single-gate (double-gate) configuration. (a) and (b): on-site Hubbard parameters $V_{00}$; (c) and (d): nearest neighbour Hubbard parameters $V_{01}$; (e) and (f): long-ranged corrected on-site Hubbard parameters $U^\ast$. The vertical lines correspond to the moir\'e unit cell length for each twist angle.}
\label{U_Us_THETA}
\end{figure*}

Fig.~\ref{U_Us_THETA} shows the variation of the Hubbard parameters as a function of the separation $\xi$ of the gate(s) from the tBLG for a range of twist angles close to the magic angle. The panels on the left-hand (right-hand) side correspond to the single-gate (double-gate) device configuration. For the on-site Hubbard parameter $V_{00}$ [Figs.~\ref{U_Us_THETA}(a) and ~\ref{U_Us_THETA}(b)], it is known that they scale linearly with the twist angle~\cite{PHD_1,NAT_I}. All twist angles, therefore, exhibit a similar dependence on the separation to the gate, but with the magnitude of $V_{00}$ scaled according to the twist angle. 

\begin{figure*}[h]
\begin{subfigure}{0.49\textwidth}
  \centering
  \includegraphics[width=0.9\linewidth]{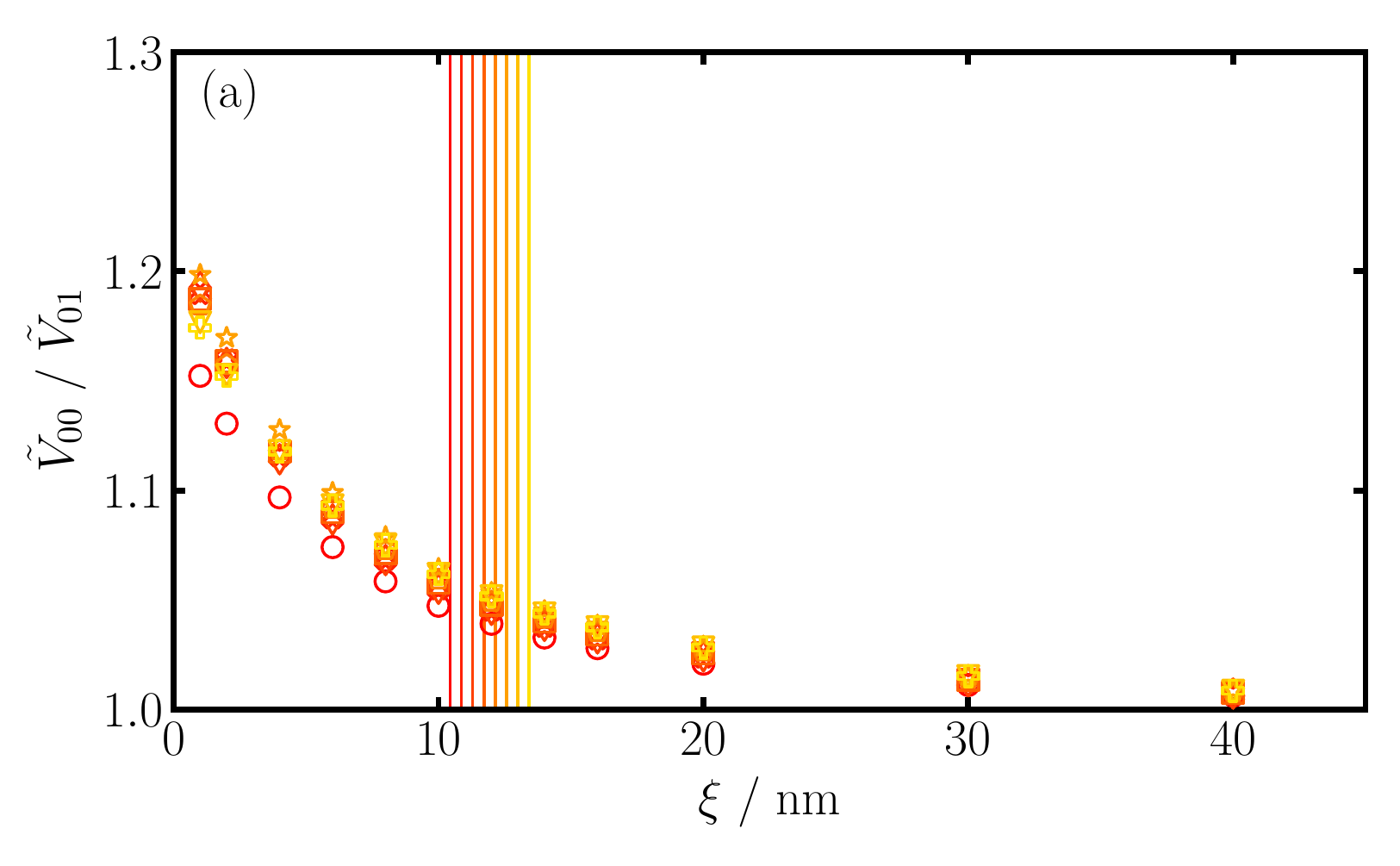}
\end{subfigure}
\begin{subfigure}{0.49\textwidth}
  \centering
  \includegraphics[width=0.9\linewidth]{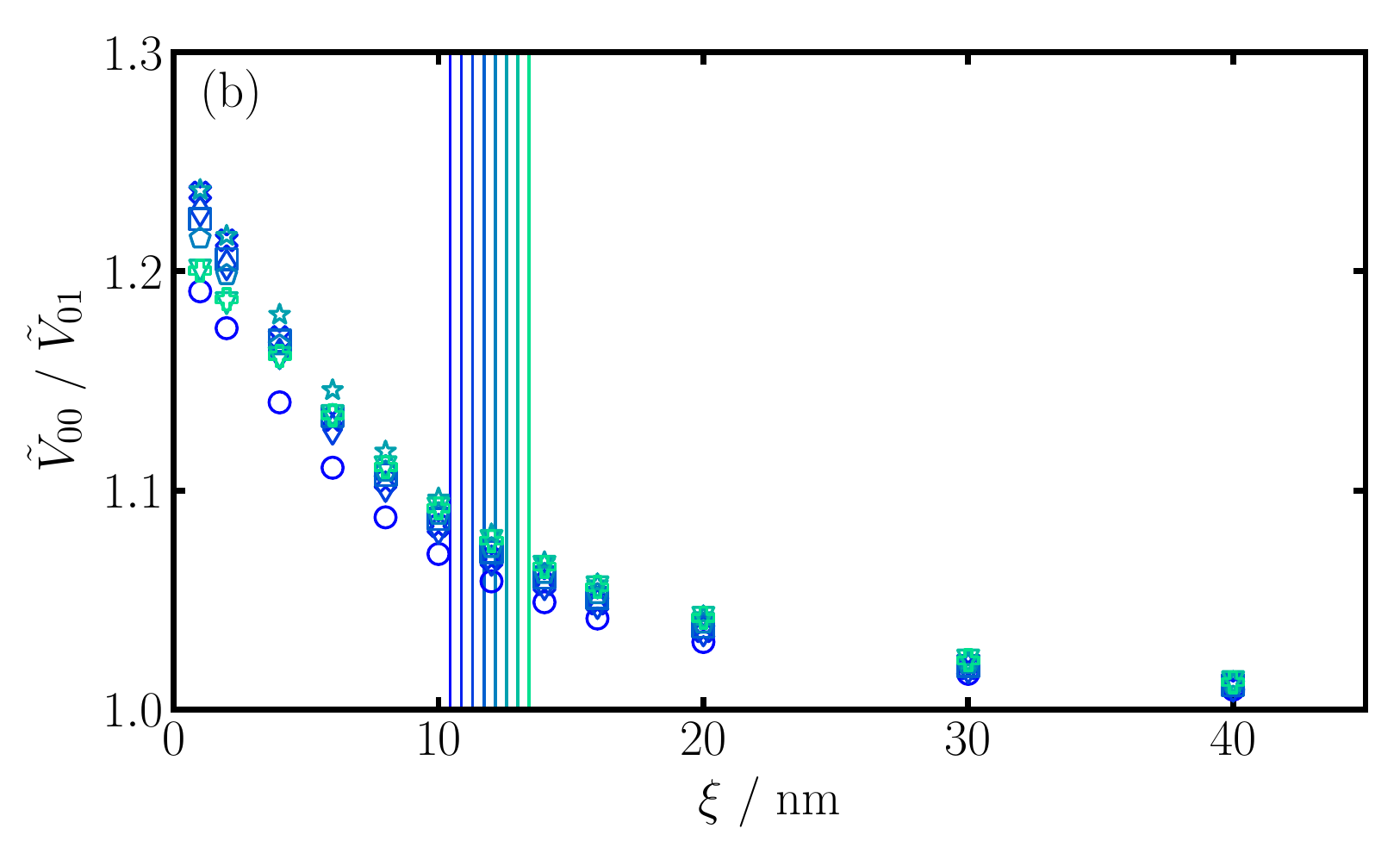}
\end{subfigure}
\caption{Ratio of on-site Hubbard parameter [$\tilde{V}_{00} = V_{00}(\xi)/V_{00}(\infty)$] to the nearest neighbour Hubbard parameter [$\tilde{V}_{01} = V_{01}(\xi)/V_{01}(\infty)$], both of which are rescaled to their Coulomb limit value, as a function of gate separation for the studied twist angles (same color and symbol coding as Fig.~\ref{U_Us_THETA}). (a) Single-gate device configuration; (b) double-gate device configuration.}
\label{V00tV01}
\end{figure*}

\begin{figure*}[h]
\begin{subfigure}{0.49\textwidth}
  \centering
  \includegraphics[width=0.9\linewidth]{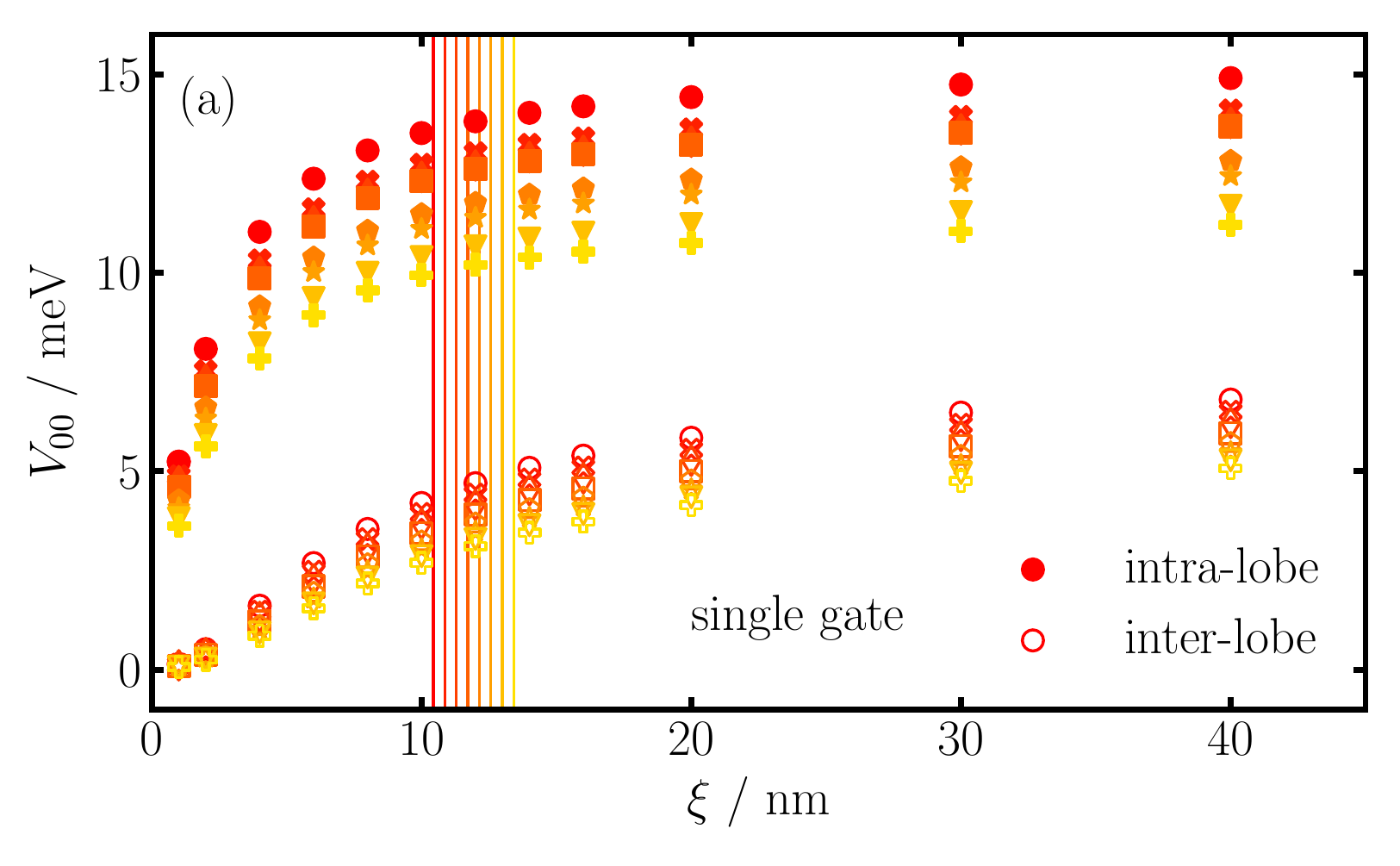}
\end{subfigure}
\begin{subfigure}{0.49\textwidth}
  \centering
  \includegraphics[width=0.9\linewidth]{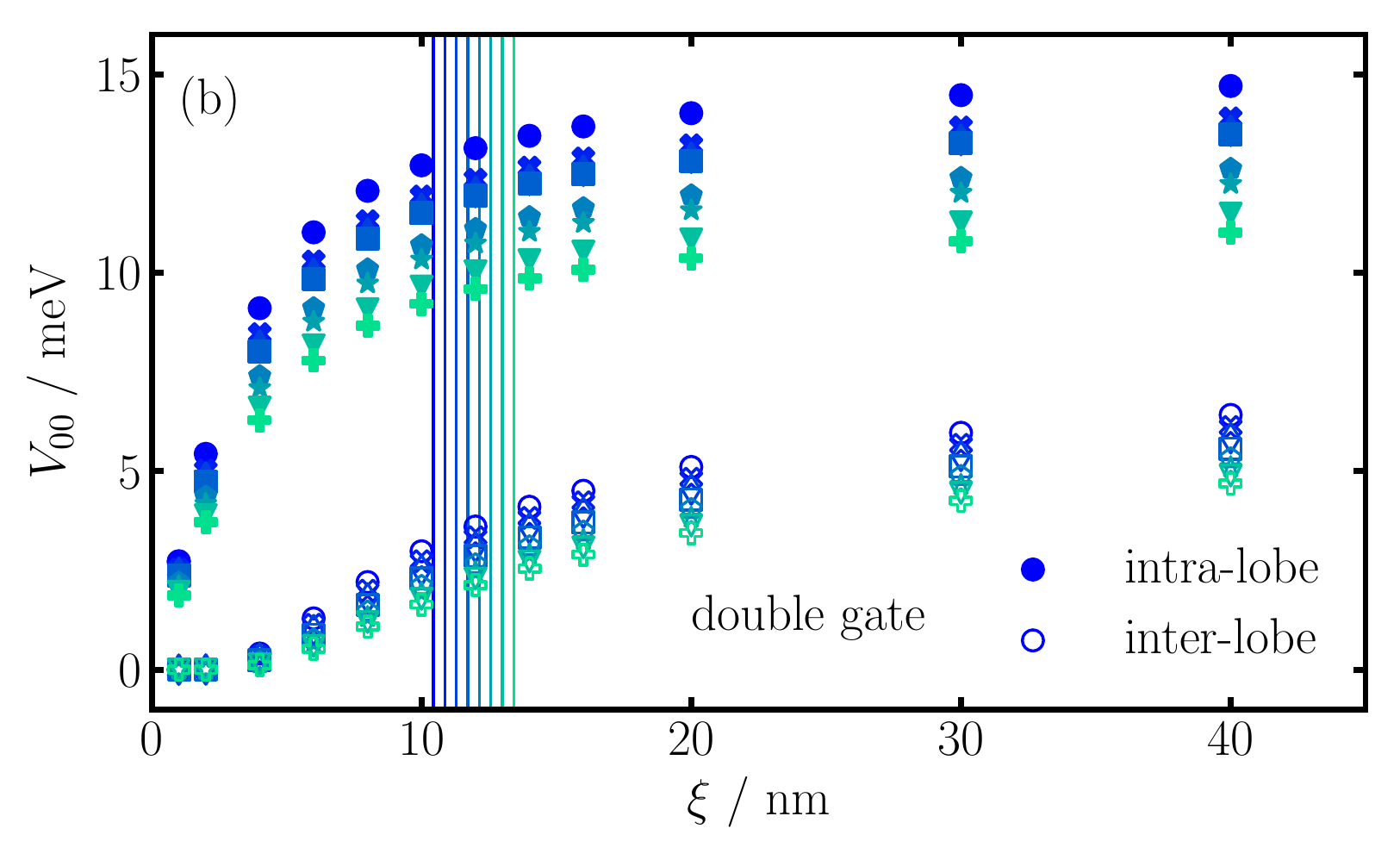}
\end{subfigure}
\begin{subfigure}{0.49\textwidth}
  \centering
  \includegraphics[width=0.9\linewidth]{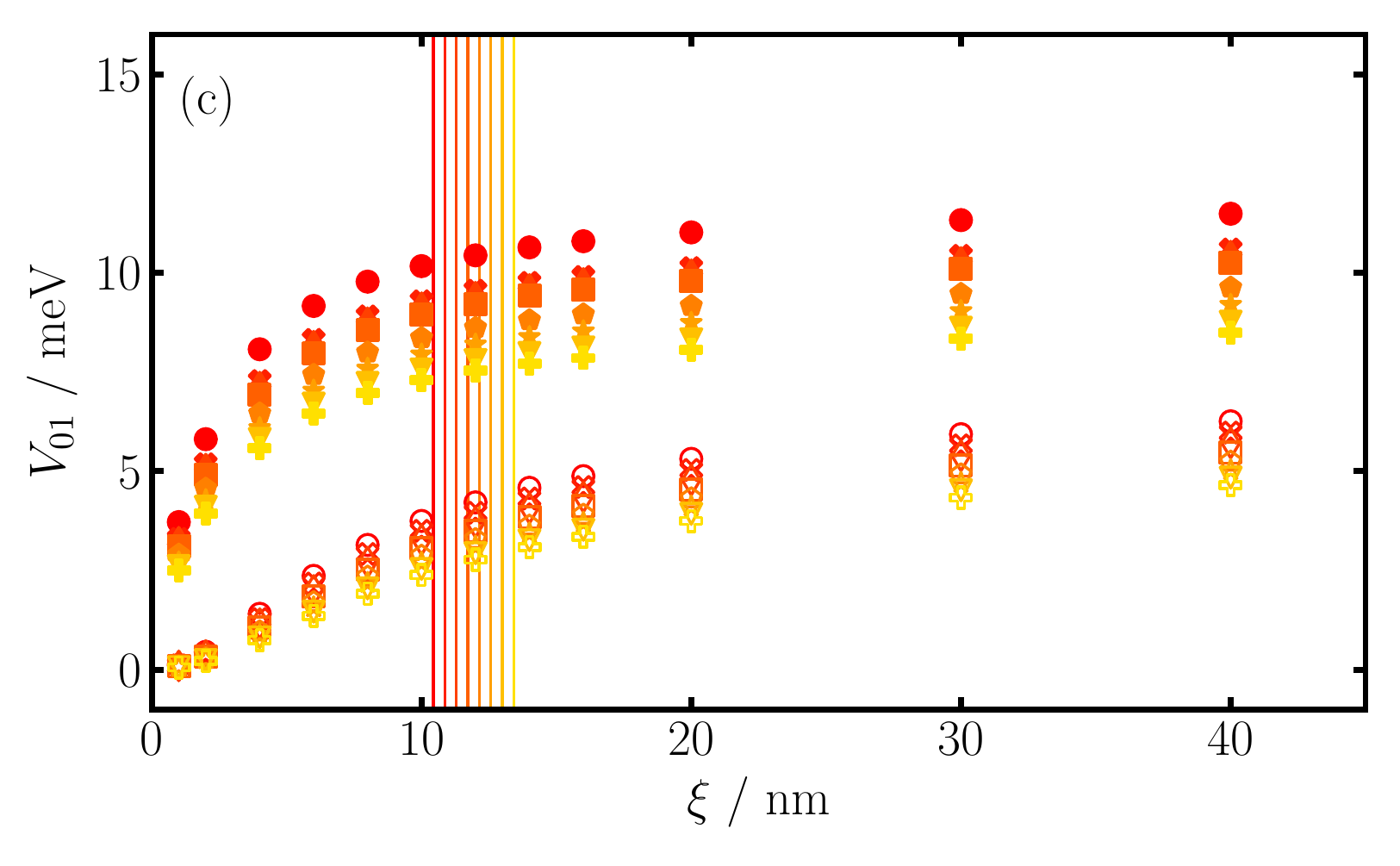}
\end{subfigure}
\begin{subfigure}{0.49\textwidth}
  \centering
  \includegraphics[width=0.9\linewidth]{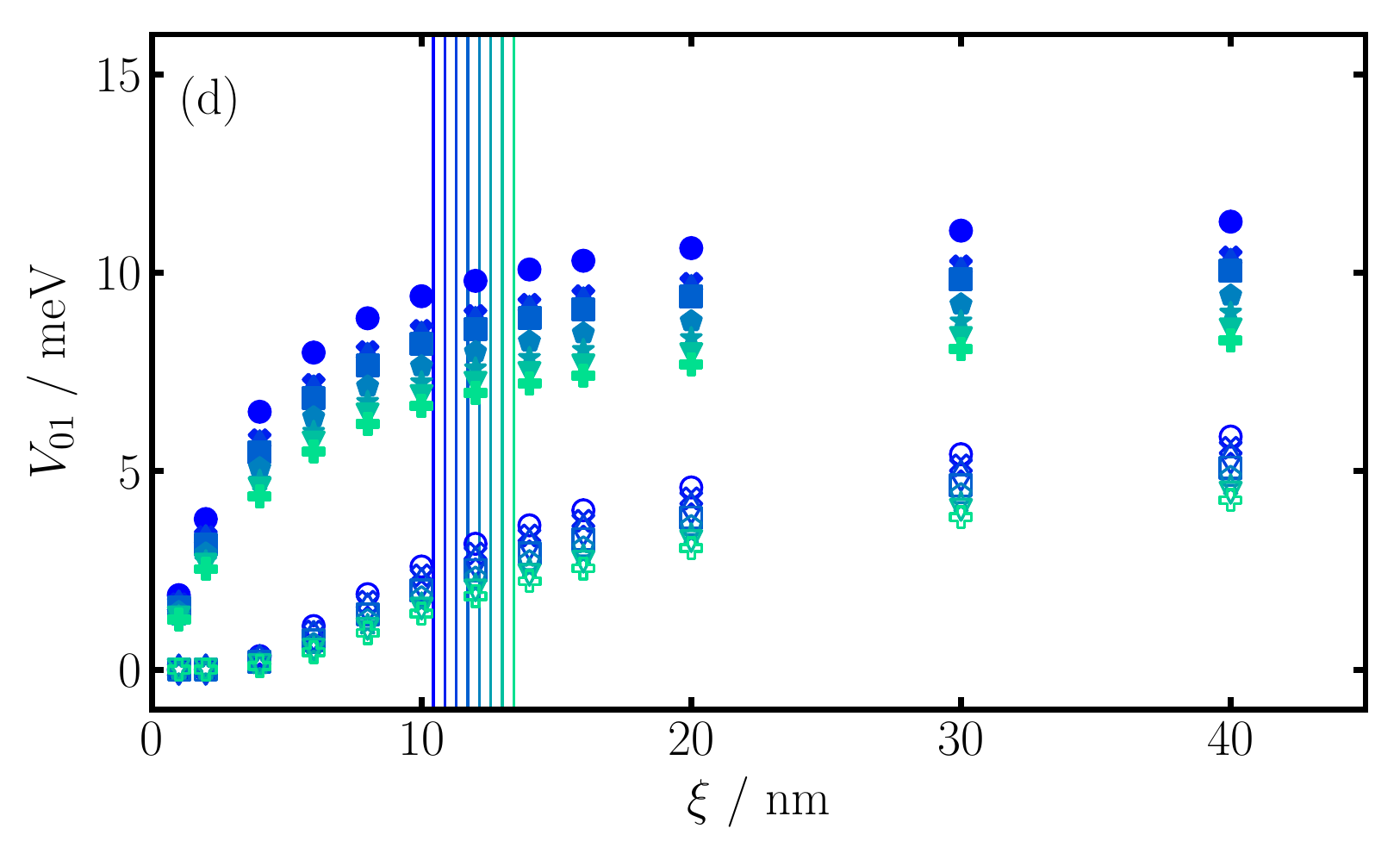}
\end{subfigure}
\caption{Intra-lobe (closed symbols) and inter-lobe (open symbols) contributions to the on-site Hubbard parameter $V_{00}$ [(a) and (b)] and nearest-neighbour Hubbard parameter $V_{01}$ [(c) and (d)] as a function of gate separation $\xi$ for the studied twist angles (same colour and symbol coding for the different twist angles as Fig.~\ref{U_Us_THETA}). The panels in the left-hand (right-hand) column are for the single-gate (double-gate) configuration. The vertical lines correspond to the moir\'e unit cell length for each twist angle.}
\label{Inter_Intra}
\end{figure*}

%In Figs.~\ref{U_Us_THETA}(c) and \ref{U_Us_THETA}(d) the next nearest Hubbard parameter for, respectively, the single gate and double gate device is shown as a function of separation to the gate for the indicated twist angles. As mentioned in the main text, these have a very similar dependence on the gate separation as the on-site Hubbard parameters. This can be rationalised from the fact that the overlap of the Wannier functions for the next nearest interaction is very similar to the interaction of the on-site Hubbard parameter. More specifically, the on-site Hubbard interaction has all three lobes of the Wannier functions overlapping, and the next nearest Hubbard interaction has two lobes of the Wannier function overlapping.
%aam:
%\textcolor{red}{[AAM: in the main text we say that $V_{01}$ decreases more rapidly than $V_{00}$ and then point to this Appendix. So I guess we should quantify the rapidity of the decrease and comment on it here. From the plots it is difficult to see by eye whether the statement is obviously true or not.]}

The nearest neighbour Hubbard parameter $V_{01}$ [Figs.~\ref{U_Us_THETA}(c) and \ref{U_Us_THETA}(d)] and, therefore, also the long-ranged corrected on-site Hubbard parameter $U^{\ast}=V_{00}-V_{01}$ [Figs.~\ref{U_Us_THETA}(e) and \ref{U_Us_THETA}(f)] have a similar dependence on the gate separation as the on-site Hubbard parameter $V_{00}$. This similarity can be understood from the three-lobe structure of the Wannier functions~\cite{MLWO,SMLWF,PHD_1,PHD_2}: the predominant contribution to the Hubbard parameters comes from the overlap of lobes that are centred on the same moir\'{e} lattice site (``intra-lobe” contributions) plus contributions from the overlap of lobes centred on neighbouring moir\'{e} lattice sites (``inter-lobe’’ contributions)~\cite{MLWO}. In the case of the on-site parameter $V_{00}$, there are three intra-lobe and six first-nearest neighbour inter-lobe interactions, whilst for the nearest neighbour parameter $V_{01}$, there are two intra-lobe interactions, six first-nearest neighbour inter-lobe interactions and one second nearest-neighbour inter-lobe interaction. 

To make the larger reduction of $V_{01}$ compare to $V_{00}$ more explicit, in Fig.~\ref{V00tV01} we display the on-site Hubbard parameter in units of the on-site Hubbard parameter in the Coulomb limit, $\tilde{V}_{00} = V_{00}(\xi)/V_{00}(\infty)$, over the next nearest Hubbard parameter in units of the next nearest Hubbard parameter in the Coulomb limit, $\tilde{V}_{01} = V_{01}(\xi)/V_{01}(\infty)$; again the subfigures are as a function of distance to the gate and all the studied twist angles are shown, with (a) corresponding to the single gate device and (b) to the double gate device. It is evident from this plot that $\tilde{V}_{00}/\tilde{V}_{01}$ increases as the distance to the gate is reduced. Therefore, $\tilde{V}_{01}$ (which also means $V_{01}$) is decreasing more than $\tilde{V}_{00}$ (and therefore $V_{01}$). 

Fig.~\ref{Inter_Intra} shows the inter- and intra-lobe contributions to $V_{00}$ and $V_{01}$ for both device configurations [(a) and (c): single-gate; (b) and (d): double-gate]. These contributions were calculated from the method suggested in Ref.~\citenum{MLWO}, but where the image-charge potential was used in place of the Coulomb potential~\cite{PHD_1}. As can be seen, the intra-lobe contributions are larger than the inter-lobe contributions for all cases~\cite{PHD_1}. As the distance to the gate is reduced (for separations around the moir\'e length scale), the inter-lobe contributions are suppressed more than the intra-lobe contributions. At separations much smaller than the moir\'e length scale, the intra-lobe contributions are significantly truncated.

\renewcommand{\theequation}{B\arabic{equation}}
\renewcommand{\thefigure}{B\arabic{figure}}
\setcounter{figure}{0} 
\setcounter{equation}{0} 

\section{Critical gate separation for the double-gate configuration}
\label{AC}

\begin{figure*}[h]
\includegraphics[width=1\linewidth]{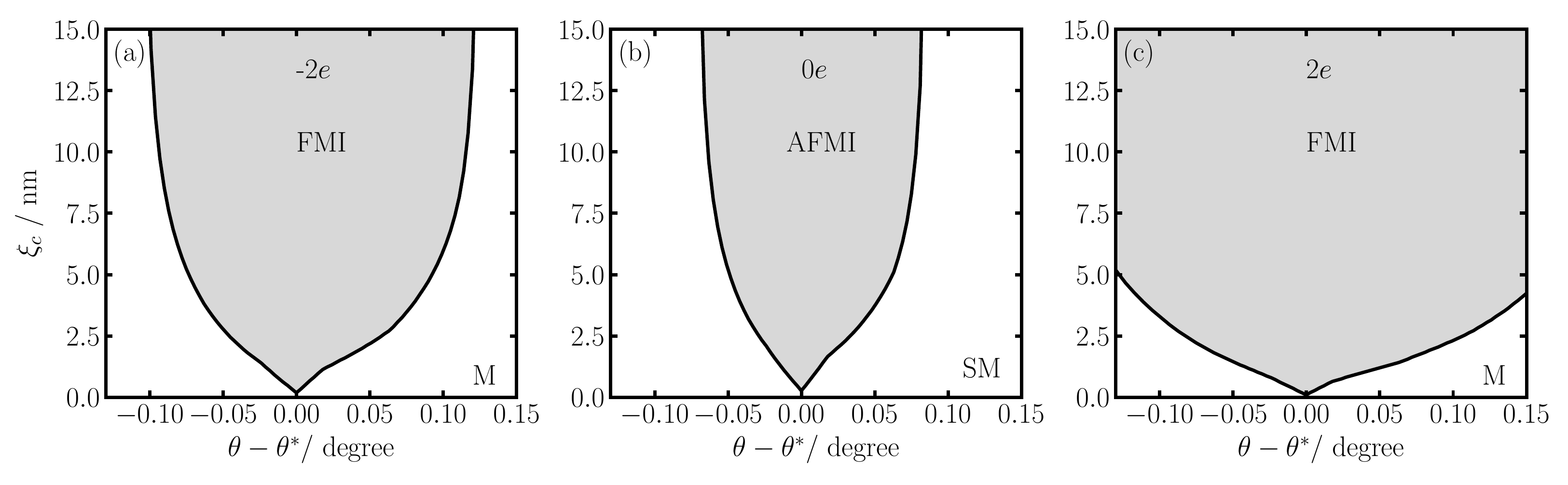}
\caption{Critical value of gate separation $\xi_{\rm c}$ as a function of twist angle from the magic angle for two additional electrons per moir\'e unit cell ($-2e$, left panel), charge neutrality ($0e$, middle panel) and two additional holes per moir\'e unit cell ($2e$, right panel), for a device in the double-gate configuration at a temperature $T\approx 0.3~K$. In the grey regions, FMI denotes a ferromagnetic insulator and AFMI denotes an anti-ferromagnetic insulator; while in the white regions, M denotes metal and SM denotes semi-metal.}
\label{Xic_WMG}
\end{figure*}

In Fig.~\ref{Xic_WMG} we show the critical gate separation for the double gate as a function of twist angle from the magic angle. The plots are qualitatively similar to those of the single-gate configuration (Fig.~\ref{Xic_SMG}). These plots were made by interpolating $U^*(\xi,\theta)$ linearly in $\theta$ and with cubic splines in $\xi$. We also fitted $t(\theta)=\Delta(\theta)/6$, where $\Delta$ is the bandwidth of the flat bands from our tight-binding model, with a cubic spline as a function of $\theta$. The roots of $U^*(\xi,\theta) - t(\theta)U/t_{\rm G}$ were found as a function of $\theta$ and $\xi$, where $U$ is the critical value of the interaction from Ref.~\citenum{LK_CH} (which is different for each value of the doping), and $t_{\rm G}$ is the hopping parameter of graphene.

\newpage

\renewcommand{\theequation}{C\arabic{equation}}
\renewcommand{\thefigure}{C\arabic{figure}}
\setcounter{figure}{0} 
\setcounter{equation}{0}

\section{Critical gate separation phase diagram at higher temperature}
\label{AD}

\begin{figure*}[h]
\includegraphics[width=1\linewidth]{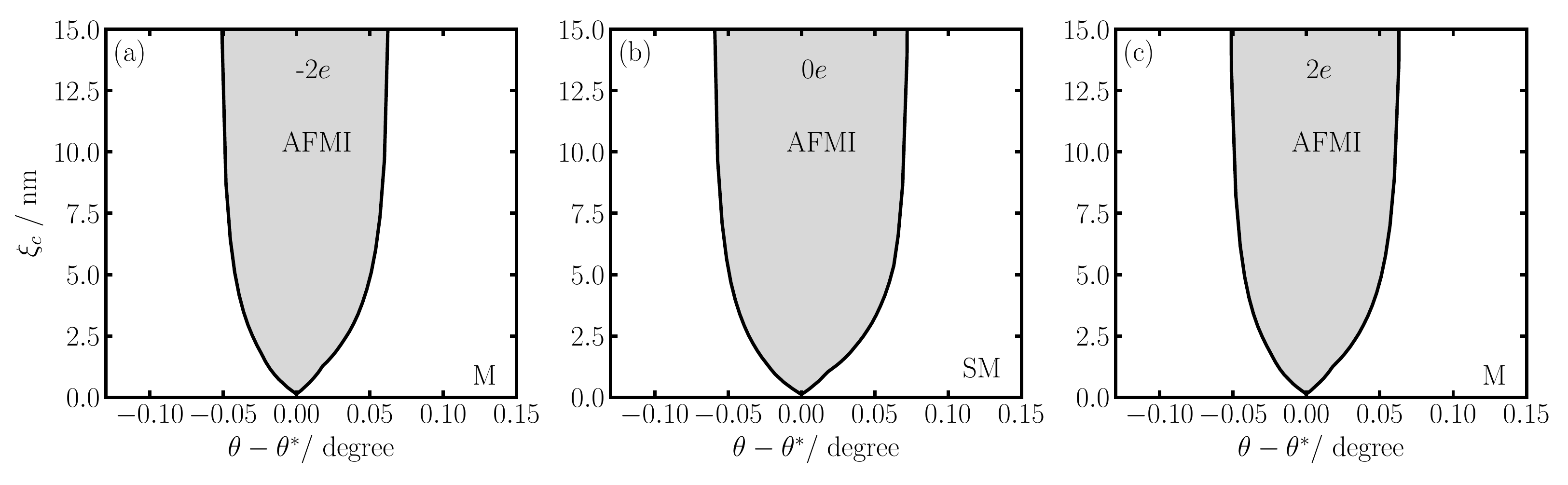}
\caption{Critical value of gate separation as a function of twist angle from the magic angle for two additional electrons per moir\'e unit cell ($-2e$, left panel), charge neutrality ($0e$, middle panel) and two additional holes per moir\'e unit cell ($2e$, right panel), for a device in the single-gate configuration at a temperature $T\approx 5~K$. In the grey regions tBLG is an anti-ferromagnetic insulator (AFMI); while in the white regions, M denotes metal and SM denotes semi-metal.}
\label{Xic_SMG_HT}
\end{figure*}

\begin{figure*}[h]
\includegraphics[width=1\linewidth]{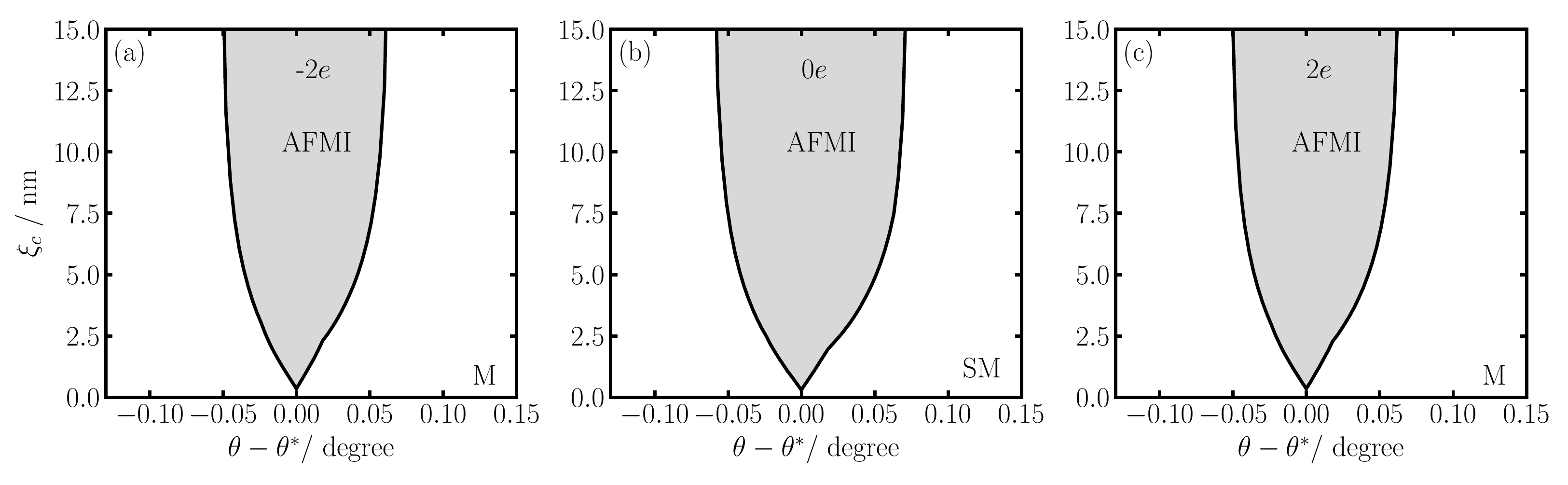}
\caption{Critical value of gate separation as a function of twist angle from the magic angle for two additional electrons per moir\'e unit cell ($-2e$, left panel), charge neutrality ($0e$, middle panel) and two additional holes per moir\'e unit cell ($2e$, right panel), for a device in the double-gate configuration at a temperature $T\approx 5~K$. In the grey regions tBLG is an anti-ferromagnetic insulator (AFMI); while in the white regions, M denotes metal and SM denotes semi-metal.}
\label{Xic_WMG_HT}
\end{figure*}

\end{document}